\documentclass[twocolumn,showpacs, prb]{revtex4}
\usepackage{graphicx}  
\usepackage{amssymb} 
\usepackage{amsmath}
\newcommand{\scc}{\mathbf{s}}
\newcommand{\sccd}{\tilde{\mathbf{s}}}
\newcommand{\sii}{\mathbf{S}}
\newcommand{\siid}{\tilde{\mathbf{S}}}
\newcommand{\jsp}{\jmath^{s}}
\newcommand{\jen}{\jmath^{\epsilon}}
\newcommand{\oprr}{\hat{O}}
\newcommand{\seff}{\mathcal{S}}
\newcommand{\seffv}{\mathcal{S}}
\newcommand{\jeff}{\tilde{J}}
\newcommand{\jgen}{\mathcal{J}}
\newcounter{conc}
\newcommand{\prconc}{\roman{conc}}
\begin{document}
\title{Transport properties of 
a spin-$\frac{1}{2}$ Heisenberg chain with an embedded spin-$S$ impurity}
\author{A. Metavitsiadis}
\affiliation{Department of Physics, University of Crete and
Foundation for Research and Technology-Hellas, P.O. Box 2208, GR-71003
Heraklion, Greece}
\date{\today}
\begin{abstract}
The finite temperature transport properties of a \mbox{spin-$\frac{1}{2}$} 
anisotropic  Heisenberg chain with an  embedded  \mbox{spin-$S$} impurity 
are studied.  Using primarily numerical diagonalization techniques, we study the 
dependence of the dynamical spin and  thermal conductivities on  
the lattice size, the magnitude of the impurity spin, 
the  host-impurity coupling, the easy axis anisotropy,  
as well as the dependence on temperature. 
Particularly for the temperature dependence, we discuss the screening 
of the impurity by the chain eventually leading  to the cutting or  
healing of the host chain. Numerical results are supported by  analytical 
arguments obtained  in the strong  host-impurity coupling regime.
\end{abstract}
\maketitle
\section{Introduction}
\par
The unconventional thermal transport properties of low dimensional 
quantum magnets have drawn the attention of the condensed matter 
society.\cite{zotos-prelovsek, 
PhysRevLett.89.156603, 
PhysRevB.66.140406,PhysRevB.67.134426, PhysRevB.67.224410,  
PhysRevB.67.064410,PhysRevB.71.184415, PhysRevB.68.104401, 
PhysRevLett.94.087201, PhysRevLett.96.067202, 
PhysRevB.79.024425, sologubenko-2007, hess-2007, nov-Otter2009796, 
JPSJ.71.2485} %
In particular, the transport properties of spin chain materials  
are described by the one-dimensional (1D) 
\mbox{spin-$\frac{1}{2}$} Heisenberg model, where the 
large  exchange coupling along the 
chain\cite{PhysRevLett.73.332, PhysRevB.53.5116, PhysRevLett.76.3212} %
accounts for an extraordinary high and anisotropic thermal 
conductivity.\cite{PhysRevB.64.054412} %
Actually, the pure anisotropic Heisenberg model (AHM) was shown to 
exhibit ballistic  thermal  transport at any temperature $T$,  attributed
to its  integrability.\cite{PhysRevB.55.11029, 0305-4470-35-9-307, 0305-4470-36-46-006} %
In accord with the theoretical predictions, evidence of 
ballistic thermal transport in real materials has been  consolidated lately using 
samples of very high purity.\cite{PhysRevB.81.020405} %
However, defects may have a dramatic effect on the transport properties
of these systems due to the reduced dimensionality.
\par
In this work, we try to shed light on the  effect of a single, \mbox{spin-$S$} 
magnetic impurity, on the transport properties of a \mbox{spin-$\frac{1}{2}$} 
Heisenberg chain. We address the behavior of the system on various problem 
parameters at high and intermediate $T$,  while we support our numerical 
results  with analytical arguments. 
It is well known that magnetic impurities get 
screened by the chain at low $T$ forming with their neighbors an effective 
impurity of different 
spin.\cite{PhysRevB.46.10866, PhysRevLett.86.516, PhysRevB.64.092410} %
Notwithstanding, as we show in this work, conspicuous excitations, 
which cannot be described by the effective impurity picture,  
survive at low $T$; this holds for a single impurity in a finite system or 
plausibly for a finite concentration of impurities in the thermodynamic limit. %
Furthermore, the screening of the impurity triggers 
Kondo-type, many body,  phenomena below a crossover 
temperature, and the chain becomes perfectly (insulating)transmitting  at $T=0$
for (anti)ferromagnetic  easy axis anisotropy  
$\Delta$.\cite{PhysRevB.58.5529, PhysRevB.81.205101} %
This is an analog of the behavior of a Luttinger liquid 
in the presence of a non-magnetic impurity  where the chain is (cut)healed for 
(repulsive)attractive interactions.\cite{PhysRevLett.68.1220} %
We seek evidence of the cutting-healing behavior of the chain  
using the  thermal conductivity  as a probe, which is ideal for this purpose 
since the  only energy current relaxation mechanism  is induced by the impurity. 
\par
From the above it is obvious that the role of the  thermal conductivity in these 
systems is exceptional and twofold. Not only is the thermal conductivity 
a very useful theoretical tool but it is of high experimental 
and technological interests as well. Thus, our results could serve 
as  qualitative guidelines for forthcoming thermal conductivity measurements 
in doped cuprates, e.g., $\mathrm{Ni}$ doping in the renowned 
 \mbox{$\mathrm{SrCuO}_{2}$}  and   
 \mbox{$\mathrm{Sr}_{2}\mathrm{CuO}_{3} $} cuprates. 
Furthermore,  magnetic impurities in highly heat conducting 
materials could function as potential switching mechanisms  
enabling tunable heat transport and the emergence of 
numerous technological applications.  
Therefore, it is vital  to understand  the fundamental 
properties of  a prototype model in the presence of magnetic impurities 
before  trying to exploit the transport properties of these truly  unique systems. 
\section{Model}
\begin{figure}[b!]
\centering
\includegraphics[scale=0.24]{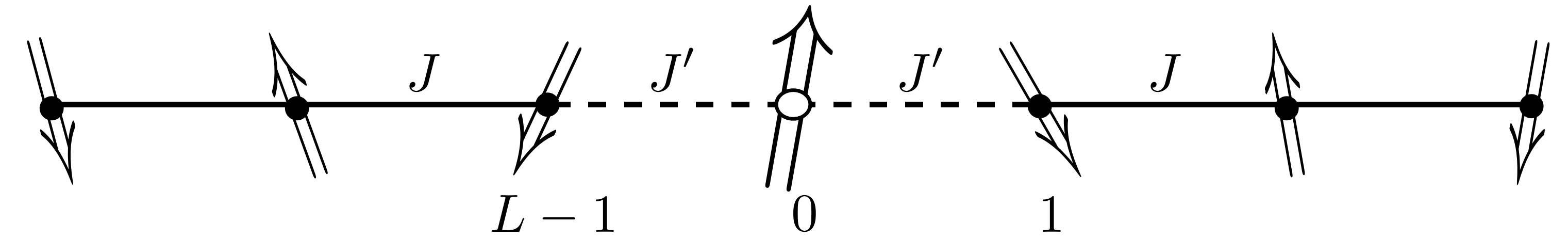}
\caption{Spin-$S$ impurity embedded in a spin-$\frac{1}{2}$ chain. 
}\label{fig-1} 
\end{figure}
\par
In a previous work,\cite{PhysRevB.81.205101} the thermal transport properties 
of a \mbox{spin-$\frac{1}{2}$} Heisenberg chain was studied in the presence 
of a spin-$S$ magnetic impurity located out of the chain. In addition,  
a uniform chain with two consecutive weak links was discussed, 
which can be considered as a special case of a spin-$\frac{1}{2}$ impurity 
embedded in the chain. In the present 
work, we deal with the generic case of a spin-$S$ magnetic impurity embedded 
in the chain, Fig.\ \ref{fig-1}. Many of the conclusions and the basic ideas reached 
in Ref.\cite{PhysRevB.81.205101} can be applied in the present work, as well,  
due to the single impurity character of the problem. However, significant 
differences in the physical results arise due to the different geometries of the 
problems. A brief discussion and comparison between the two models is given 
at the end of the manuscript. 
\par
The Hamiltonian of an  AHM on a ring of $L$ sites, 
where  a \mbox{spin-$\frac{1}{2}$} operator, say the $\mathbf{s}_{0}$, 
is substituted by another one of different spin, say $\mathbf{S}$, 
is given by 
\begin{equation}\label{hamiltonian}
H=\sum_{l=1}^{L-2} h_{l}+J'~\big(\scc_{1}+\scc_{L-1}\big)\cdot \siid\:, \quad
\big(h_{l}=J \scc_{l}\cdot\sccd_{l+1}\big)\:,
\end{equation}
see  Fig.\ \ref{fig-1} as well. In Eq.\ (\ref{hamiltonian}), 
$J$  is the antiferromagnetic exchange coupling---with energy units---between  
\mbox{spin-$\frac{1}{2}$} operators,  $J'$ is the host-impurity coupling,  
while the tilded operators $\sccd$, $\siid$ denote  the vector operators 
$(s^{x},s^{y},\Delta s^{z})$,  $(S^{x},S^{y},\Delta S^{z})$, respectively.   
In addition, we work in a system of units where the lattice constant $a$,  
and the Planck and Boltzmann  constants are  $a,\hslash,k_{B}=1$; yet numerical 
results are presented for $J=1$.
\par
The spin $\jsp$, energy $\jen$  current operators 
are determined by the continuity  equation 
$\partial_{t}{\oprr}_{l}+\nabla\jgen_{l}=0$. Taking 
$\oprr_{l}$ to be  the local spin, energy operators and 
$\jgen_{l}$ to be the local spin, energy current operators 
respectively, we arrive at
\begin{equation}\label{spcur}
\jsp= \sum_{l=1}^{L-2} J\big(\scc_{l}\times\scc_{l+1}\big) \cdot \hat{e}_{z}+
J'\big[\big(\scc_{L-1}-\scc_{1}\big)\times\sii\big]\cdot\hat{e}_{z}\:,
\end{equation}
where $\hat{e}_{z}$ is the unit vector along the $z$-axis, and 
\begin{eqnarray}
\jen&=&\sum_{l=2}^{L-2}J^{2}\scc_{l}\cdot\big(\sccd_{l+1}\times\sccd_{l-1}\big)+
JJ'\scc_{1}\cdot\big(\sccd_{2}\times\siid\big)\nonumber\\ &+&
JJ'\scc_{L-1}\cdot\big(\siid\times\sccd_{L-2}\big)+
J'^{2}\sii\cdot\big(\sccd_{1}\times\sccd_{L-1}\big)\:. \label{encur}
\end{eqnarray}
\par 
Within linear response theory, the real part of the spin $\sigma'$, 
thermal $\kappa'$ conductivities are given 
by\cite{JPSJ.12.570,PhysRev.135.A1505,PhysRevB.73.085117}
\begin{equation}
\sigma'(\omega)=2\pi D_S \delta(\omega)+\sigma(\omega)\:,~~
\kappa'(\omega)=2\pi D_E \delta(\omega)+\kappa(\omega)\:, \nonumber
\end{equation}
where the corresponding spin $D_S$, energy $D_E$ stiffnesses denote the 
presence of ballistic transport in the system. The regular components 
$\sigma$, $\kappa$ of the corresponding spin, thermal conductivities  are 
given by
%
\begin{eqnarray}
\sigma(\omega)\hskip-0.1cm&=&\hskip-0.1cm\frac{\pi}{L}\frac{1-e^{-\beta\omega}}{\omega}
\hskip-0.15cm\sum_{\substack{n,m \\ (\epsilon_{n}\neq \epsilon_{m})}}\hskip-0.15cm
p_n|\langle n|\jsp| m\rangle|^{2}
\delta(\omega_{mn}-\omega),~~  \label{RegularConductivitys}\\
\kappa(\omega)\hskip-0.1cm&=&\hskip-0.1cm\frac{\pi\beta}{L}\frac{1-e^{-\beta\omega}}{\omega}
\hskip-0.3cm\sum_{\substack{n,m \\ (\epsilon_{n}\neq \epsilon_{m})
}}\hskip-0.2cm
p_{n} |\langle n|\jen| m\rangle|^{2}
\delta(\omega_{mn}-\omega),~~\label{RegularConductivityk}
\end{eqnarray}
where $\epsilon_n$ are the eigenvalues and 
$|n\rangle$ are the eigenstates  of the Hamiltonian (\ref{hamiltonian}), 
$p_n=\exp(-\beta\epsilon_n)/Tr \exp(-\beta H)$, 
$\omega_{mn}=\epsilon_m-\epsilon_n$, and $\beta=1/T$. %
\par
While in the pure AHM it is clear that $D_E$ is finite for any value 
of the easy axis anisotropy 
$\Delta$,\cite{PhysRevB.55.11029, 0305-4470-35-9-307, 0305-4470-36-46-006} 
there is an ongoing debate on  
the behavior of the spin stiffness.\cite{PhysRevLett.103.216602,
PhysRevB.68.134436,JPSJS.74S.181, 
PhysRevE.82.040103, PhysRevLett.88.077203, PhysRevLett.65.243, PhysRevB.53.983,
PhysRevB.59.7382, PhysRevB.58.R2921} 
Nevertheless, a single impurity renders ballistic transport 
incoherent and both $D_S$, $D_E$ vanish.\cite{PhysRevB.80.125118} %
Thus, the static component of the transport quantities is given 
by the resistive dc spin, thermal  conductivities obtained by 
the zero frequency limit of the regular components 
(\ref{RegularConductivitys}),  (\ref{RegularConductivityk}), 
viz.,  $\sigma_{dc}=\sigma(\omega\rightarrow0)$, 
$\kappa_{dc}=\kappa(\omega\rightarrow0)$.
%
%
%
\par
To  numerically study transport quantities for  systems with a  Hilbert space 
 dimension up to $\mathcal{D}\sim10^{4}$, at high temperatures,   
we employ the  exact  diagonalization (ED) technique. 
The \mbox{$\delta$-peaks} at the excitation frequencies are binned    
in windows $\delta\omega=0.01$ while we introduce an additional 
broadening  $\eta=0.03$ using the well-known Kramers-Kronig relations. 
For $\mathcal{D}\gtrsim10^{4}$, 
we use  the  microcanonical   Lanczos method\cite{PhysRevB.68.235106} 
 (MCLM)   at high temperatures and the finite 
temperature Lanczos method\cite{sic-ftlm} (FTLM) at  finite temperatures;   
yet we   keep the same additional  broadening. Lastly,  all results are obtained 
 in the  $S^{z}_{\text{total}}=0$ subsector.   
\section{Numerical results}
\subsection{Spin-1 impurity}
\par
\begin{figure}
\centering
\includegraphics[scale=0.30]{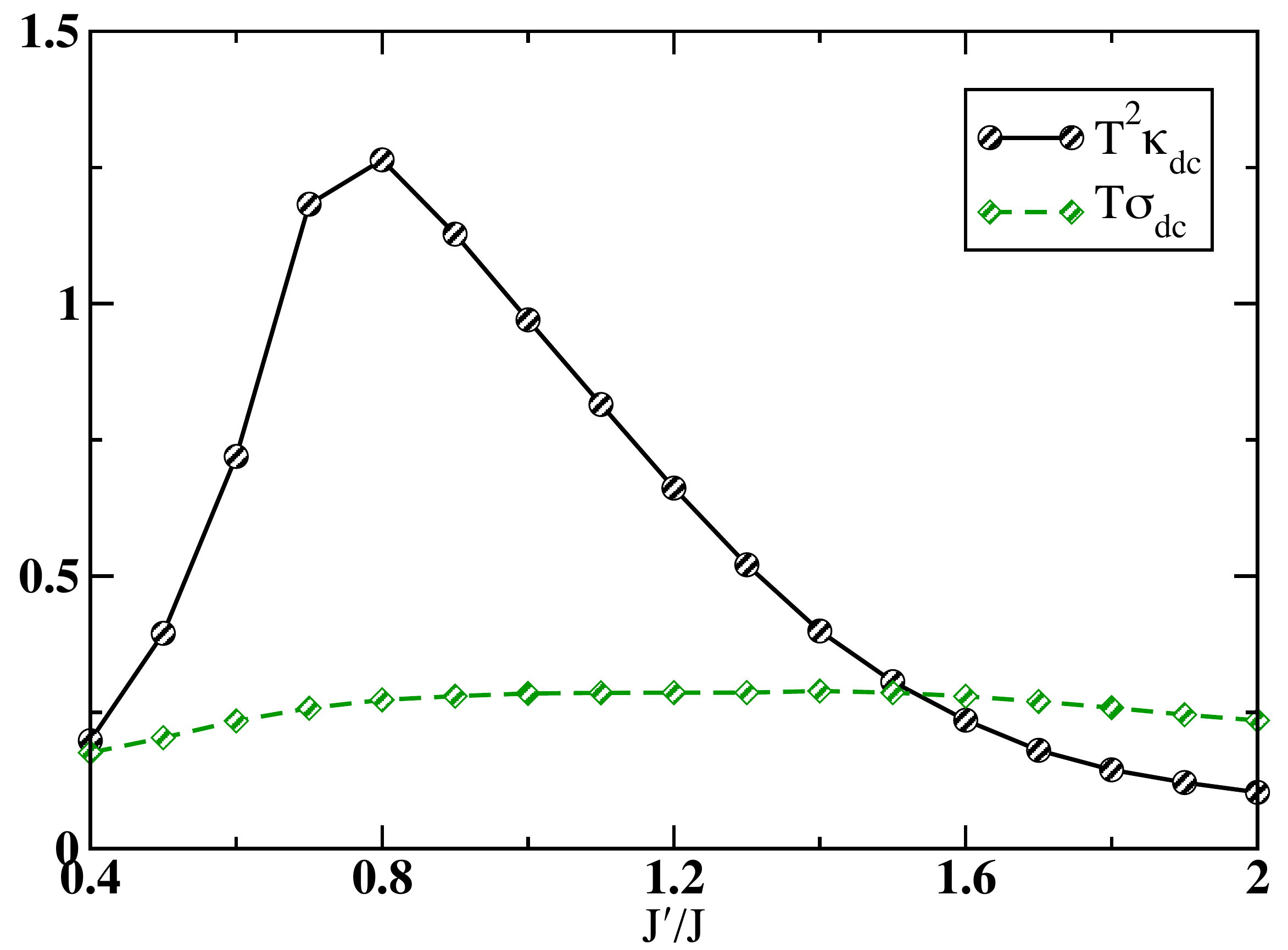}
\caption{(color online).\  
The dc value of the thermal $T^{2}\kappa_{dc}$, spin  $T\sigma_{dc}$ 
conductivity is shown as a function of the host-impurity coupling $J'/J$,  obtained 
via ED for $L=15$, $S=1$, $\Delta=1$, and  $\beta\rightarrow0$. }\label{ks-dc}
\end{figure}
\begin{figure}
\centering
\includegraphics[scale=0.30]{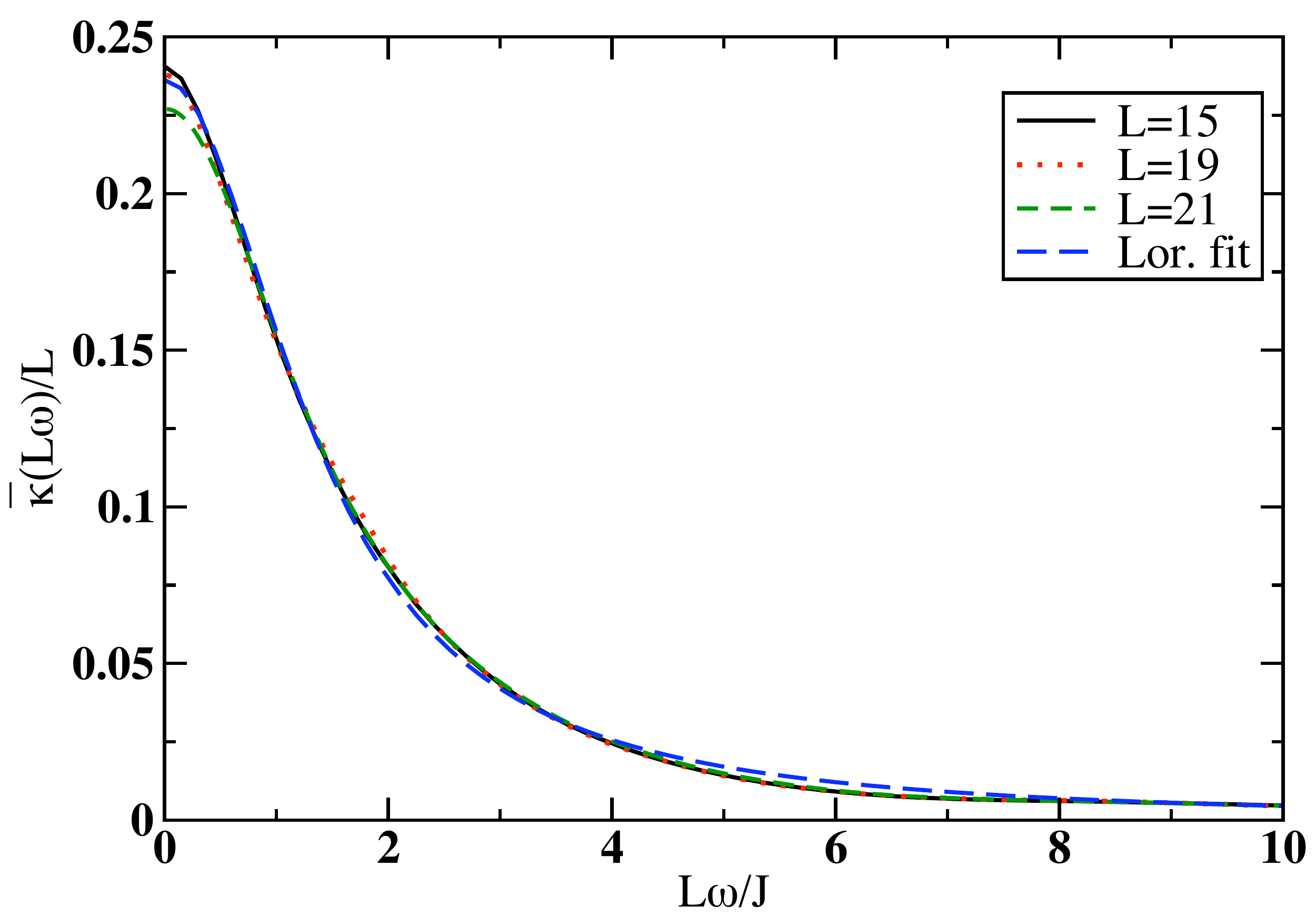}
\caption{(color online).\  
The frequency dependence of the normalized and scaled 
thermal conductivity $\bar{\kappa}(\omega L)/L$ is shown 
for $L=15,19,21$ and $J'/J=0.8$; 
in addition  a Lorentzian fit is depicted. 
Results are obtained for $S=1$, $\Delta=1$, and  $\beta\rightarrow0$. 
$L=15$ results are obtained  via ED and $L>15$ via MCLM.}\label{klor}
\end{figure}
\begin{figure}
\centering
\includegraphics[scale=0.30]{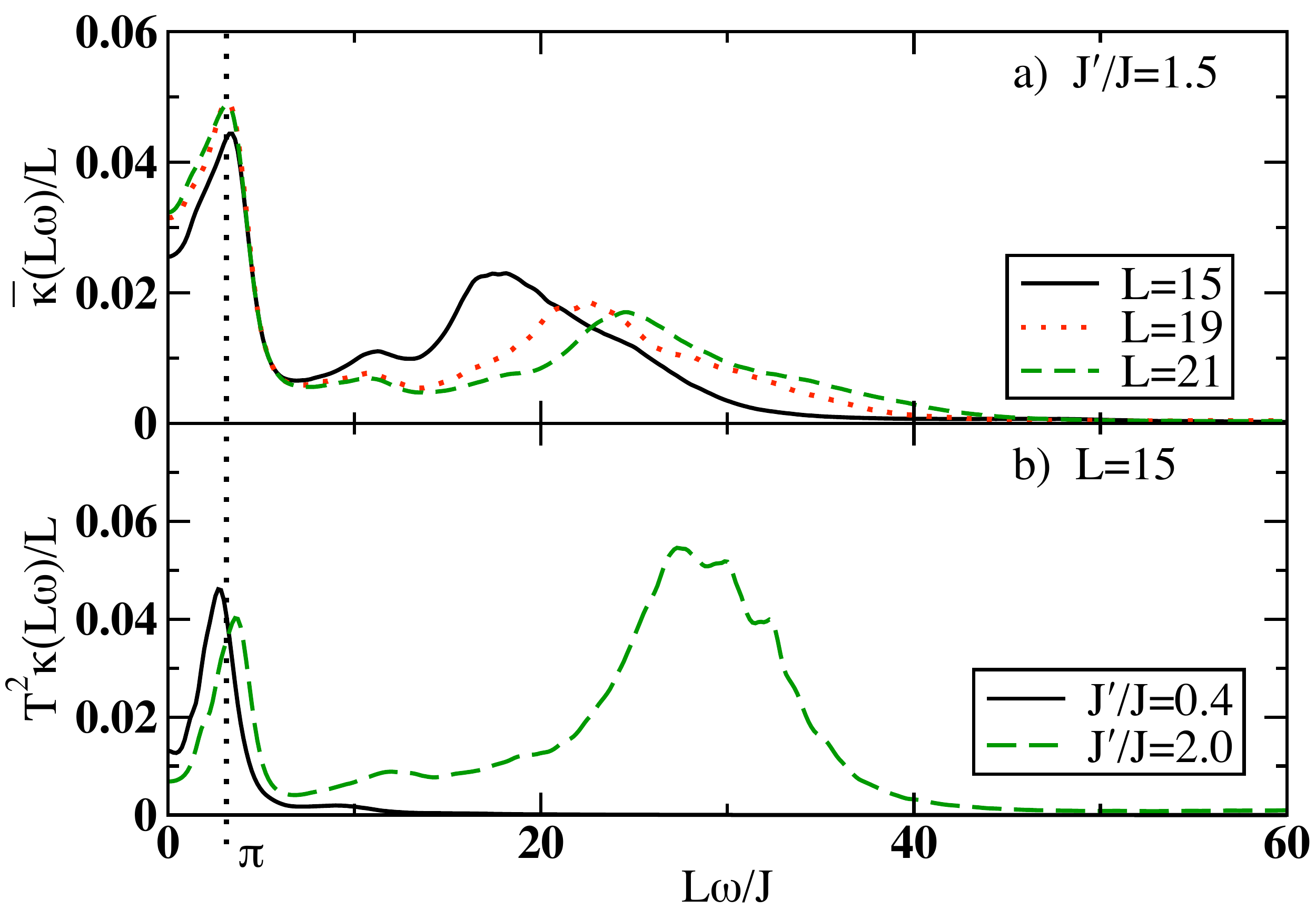}
\caption{(color online).\  
a) The frequency dependence of the normalized and scaled 
thermal conductivity $\bar{\kappa}(\omega L)/L$  is shown 
for $L=15,19,21$ and $J'/J=1.5$. 
b) The frequency dependence of $T^{2}\kappa(\omega L)/L$ 
is shown for $L=15$ and two 
host-impurity  couplings $J'/J=0.4,2.0$. %
Results are obtained for $S=1$, $\Delta=1$, and $\beta\rightarrow0$. 
$L=15$ results are obtained  via ED and $L>15$ via MCLM. 
}\label{k-ws} 
\end{figure}

First, we would like to address the $S=1$ impurity case  since this may be 
the most appealing magnetic impurity doping  for experiments. 
In Fig.\ \ref{ks-dc}, we present results for the dc spin $\sigma_{dc}$, 
thermal $\kappa_{dc}$  conductivities, while in Figs.~ \ref{klor}, and \ref{k-ws} we 
present the dynamical  
thermal conductivity of the  isotropic ($\Delta=1$) Heisenberg chain.   
A  wide range  of host-impurity 
couplings, $J'/J=0.4-2.0$, is shown at high temperatures,  $\beta\rightarrow0$,  
for various lattice sizes, $L=15-21$.   
Results for $L=15$ are obtained 
 via the ED technique, while results for $L>15$ are obtained by 
 employing the MCLM.  In order to eliminate  unimportant 
 for this  discussion prefactors  of the 
transport quantities, we plot either the normalized thermal conductivity,   
$\bar{\kappa}(\omega)=\kappa(\omega)/\kappa_{0}$ where the normalization 
$\kappa_{0}$ 
is given by $\kappa_{0}=\int \kappa(\omega) d\omega$, %
or the non-trivial at high temperatures 
$T\sigma_{dc}$, $T\sigma(\omega)$,  $T^{2}\kappa_{dc}$ and $T^{2}\kappa(\omega)$. 
Finally, for the discussion of the lattice size scaling we plot 
the scaled thermal conductivity  
$\kappa(\omega L)/L$.\cite{PhysRevB.81.205101, PhysRevB.80.125118} %
%
%
\par 
For a very weak or very strong $J'$,  a severe reduction of  $\kappa_{dc}$ 
from its maximum value, which occurs at $J'=0.8J$, is illustrated 
 in Fig.\ \ref{ks-dc}. %
On the contrary, the behavior of $\sigma_{dc}$ is qualitatively different   
from the one of $\kappa_{dc}$.  
Besides the qualitative difference, it is rather impressive that $\sigma_{dc}$ hardly 
changes in the wide range of $J'$ shown in Fig.\ \ref{ks-dc}. 
Although it is reasonable that 
$\sigma$ will be less sensitive to the effect of the  single impurity  due to the 
bulk scattering---$[H,\jsp]\neq0$ in the pure model---its rigidity is still surprising.  
\par
In Fig.\ \ref{klor},   $\bar{\kappa}(\omega L)/L$ is depicted
for  $J'=0.8J$ and a  Lorentzian fit as well,  
signifying  the Lorentzian behavior of $\kappa$, 
$\kappa(\omega)=\kappa_{dc}/[1+(\omega\tau^{\epsilon})^{2}]$ with 
$\tau^{\epsilon}$ the scattering time. 
The thermal conductivity retains  its Lorentzian shape, approximately, 
in a range of values of the host-impurity coupling,  $J'/J\simeq0.8\pm0.2$. 
As previously proposed,\cite{PhysRevB.81.205101}  a Lorentzian 
form of $\kappa(\omega)$ is an indication of a weak perturbation, 
while on the contrary a non-monotonic form  implies that the system has flown 
to  the strong perturbation regime.
The Lorentzian behavior is retained, with an almost constant $\tau^{\epsilon}$,  
for temperatures as  low as the limiting FTLM temperature for finite size systems 
$T_{fs}$ below which FTLM results are not reliable;\cite{sic-ftlm} %
we estimate $T_{fs}/J\simeq0.3$.
Although $T_{fs}$ is sufficient to reveal the cutting 
of the chain for strong perturbations,\cite{PhysRevB.46.10866} %
 for weak perturbations a lower $T_{fs}$ is  required.\cite{PhysRevB.81.205101} %
To obtain an impression of  $T_{fs}$, consider that  
for systems like the \mbox{$\mathrm{Sr}_{2}\mathrm{CuO}_{3}$} compound   
with $J/k_{B}\approx2500$ K, $T_{fs}\approx750$ K. However, 
defects may reduce $J$ of  doped samples\cite{PhysRevB.64.092410}  
bringing  $T_{fs}$  closer to  room  temperature. 
\par
For extreme values of the coupling $J'$, either strong or weak, $\kappa(\omega)$
exhibits a strongly non-monotonic behavior, 
Figs.\ \ref{k-ws}(a), and \ref{k-ws}(b), indicating  
that the system couples strongly to the impurity in both cases.  
Whereupon, the low frequency 
behavior ($L\omega/J \lesssim\pi$),
corresponding to an open-like chain,\cite{PhysRevB.77.161101} 
is similar  for a weak and a strong 
$J'$,  Fig.\ \ref{k-ws}(b). %
On the other hand, the high frequency behavior of $\kappa(\omega)$ for  
a weak and a strong $J'$ is  strikingly different due to the emergence   
of a conspicuous secondary structure at high frequencies. 
The frequency of this  structure shifts with $J'$,
indicating that its origin is 
local excitations of the impurity. Moreover,  the larger the $J'$, the more 
weight is accumulated  at this structure,
which  becomes the prevalent contribution to $\kappa(\omega)$ 
for fairly strong couplings, despite being only a $1/L$ effect. 
\subsection{Spin-$S$ impurity}
\par
The fact that the contribution of a $1/L$ effect to $\kappa(\omega)$
surpasses the bulk contribution may seem quite bizarre, however, we
can comprehend the origin of this effect from the analytical expression  
 of  the  sum rule of $\kappa(\omega)$. 
Since the energy stiffness $D_{E}$ vanishes in  the presence of impurities 
the sum rule of the thermal conductivity $\kappa'(\omega)$ is equal to the 
zeroth moment ($\kappa_{0}$) of the regular part $\kappa(\omega)$.
Starting from Eq. (\ref{RegularConductivityk}) and taking the high temperature  
limit  the sum rule of the thermal conductivity can be written  as 
 the thermodynamic 
average of the square of the energy current operator\cite{PhysRevLett.92.067202} 
\begin{equation}\label{Sumrule1}
\int_{-\infty}^{+\infty}\hskip-0.25cm \kappa(\omega)d\omega=\frac{\pi\beta^2}{L} \langle \jen \jen \rangle,~~
\text{with, }~~~
\langle \oprr \rangle = \frac{Tr e^{-\beta H} \oprr}{Tr e^{-\beta H}},   
\end{equation}
and  $\pi\beta^2 \langle \jen \jen \rangle/L=\kappa_{0}$. 
Taking the infinite temperature limit ($\beta\rightarrow0$) in the 
evaluation of the thermodynamic average, Eq.\ (\ref{Sumrule1}),  one arrives at 
\begin{equation}\label{SumRule}
\kappa_{0}=\pi\frac{1+2\Delta^{2}}{T^{2}}\left(
\frac{J^{4}}{32}\Big(1-\frac{3}{L}\Big)
+\frac{\mathcal{B}^{2}}{8L}\Big(2J^{2}
+J'^{2}\Big)\right)\:. 
\end{equation}
$\mathcal{B}^{2}=\frac{1}{3}J'^{2}S(S+1)$ is  the characteristic spin 
impurity dependence. The bulk contribution, $\sim J^{4}$, to $\kappa_{0}$ 
is equal to the impurity contribution, $\sim J'^{4}$   
(we omit  the  $\sim (JJ')^{2}$ terms), 
for a coupling 
$J'\simeq J^{*}=J\sqrt[4]{\frac{3}{4}\frac{L-3}{S(S+1)}}$. For one thing, 
for a finite system studied via ED of 
$L=13$, $S=2$ we have  $J^{*}\simeq J$. Thus, $\kappa(\omega)$  will 
exhibit resonant modes for  strong perturbations, at frequencies 
$\omega\sim J'$---at least at high  temperatures.  %
%
\begin{figure}
\centering
\includegraphics[scale=0.3]{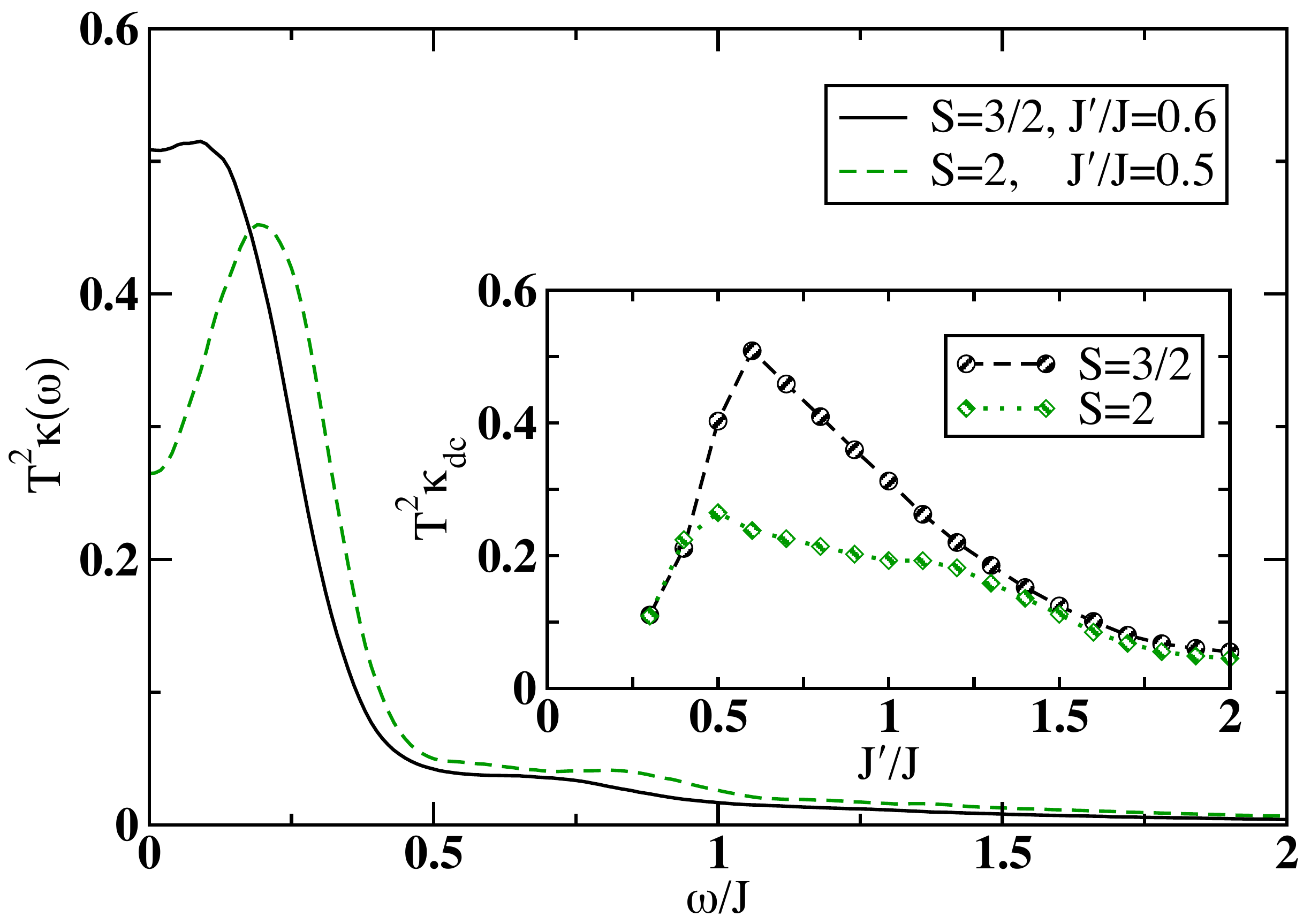}
\caption{(color online).\ The frequency dependence of the thermal conductivity 
$T^2\kappa(\omega)$ is shown for two impurities $S=3/2,2$ with 
$J'/J=0.6,0.5$ and $L=14, 13$, respectively.  
Inset: $T^2\kappa_{dc}$ as a function of the coupling $J'/J$ for $S=3/2,2$. 
Results are obtained via ED for  $\Delta=1$, $\beta\rightarrow0$.
}\label{kvrs} %
\end{figure} %
\begin{figure}
\centering
\includegraphics[scale=0.3]{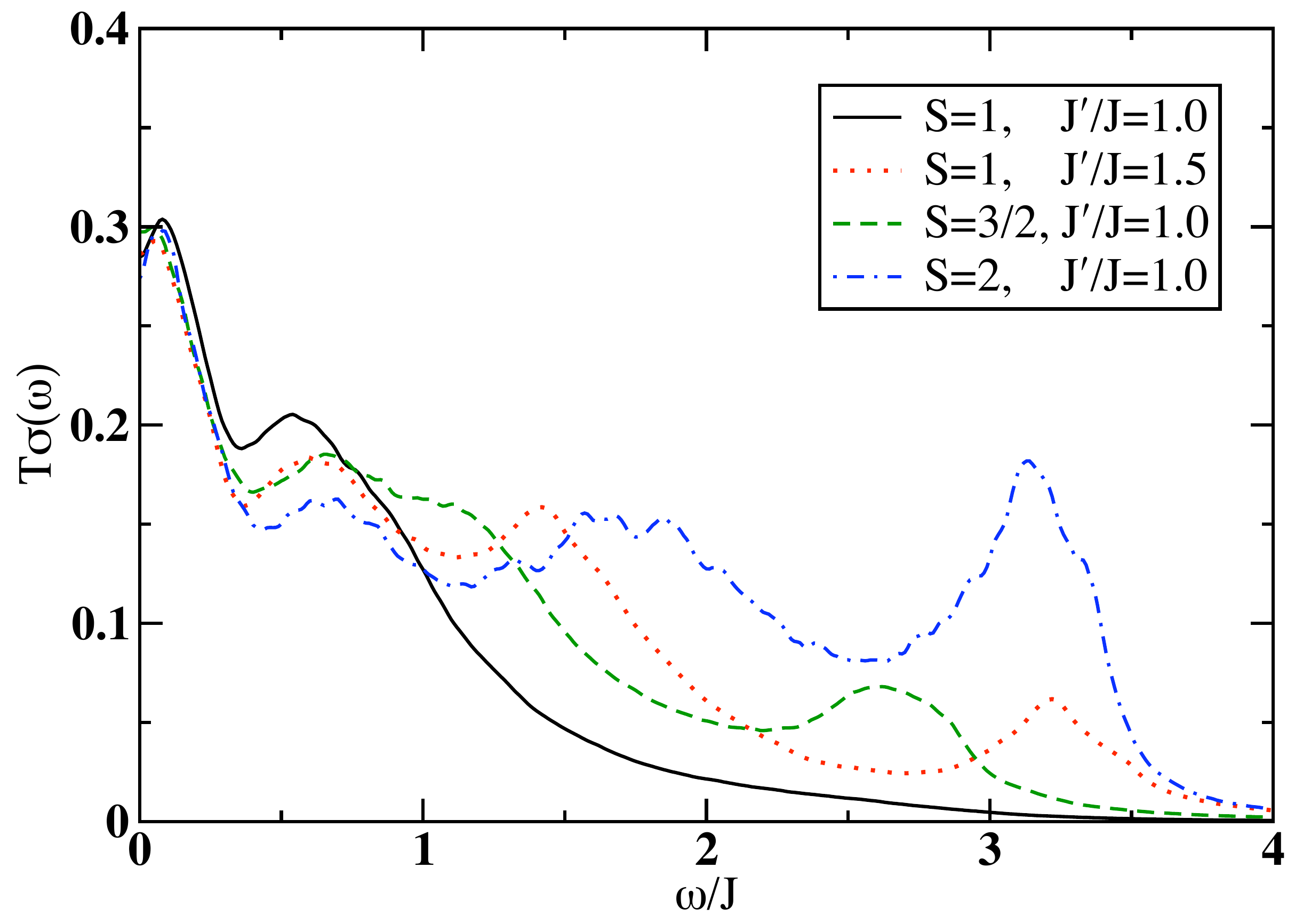}
\caption{(color online).\  
The frequency dependence of the spin conductivity $T\sigma(\omega)$  is shown 
for various perturbations: $S=1,3/2,2$ with  $J'/J=1.0$ and $L=15, 14, 13$, respectively,  as well as the case  $S=1$, $L=15$ with $J'/J=1.5$. %
Results are obtained via ED for $\Delta=1$, $\beta\rightarrow0$. 
}\label{svrs} 
\end{figure}

\par
Another important conclusion that arises from the evaluation of $J^{*}$ is 
that the larger the $S$ the smaller the $J^{*}$, since $J^{*}\sim 1/\sqrt{S}$. 
Thus, the local excitations of the impurity dominate even more easily 
over the bulk contribution for larger $S$. As a matter of fact,  we did not 
obtain a Lorentzian  $\kappa(\omega)$ for any $S>1$ by adjusting 
the host-impurity coupling, implying that any $S>1$ constitutes a strong 
perturbation for $\kappa(\omega)$ while $S=1$ seems to be quite unique.  
As far as $\sigma(\omega)$ is concerned, its high frequency regime  is 
dominated  by  the impurity local excitations, similarly to $\kappa(\omega)$,  
but its low frequency  $\sigma(\omega\rightarrow0)$ regime remains 
virtually unaffected for higher $S$. This is a striking  difference between 
$\sigma_{dc}$ and $\kappa_{dc}$
where the former shows a surprising rigidity to the influence of the 
impurity (large $S$, strong/weak $J'$) while the latter is severely reduced 
in the strong perturbation regime. %
The above arguments are summarized in Figs.\ \ref{kvrs}, and \ref{svrs} where  
the frequency dependence of the thermal, spin conductivity is shown respectively. 
In  Fig.\ \ref{kvrs},  $T^2\kappa(\omega)$ is shown for $S=3/2,2$
and $J'/J=0.6,0.5$, namely, the corresponding couplings for which $\kappa_{dc}$ is 
maximum, Fig.\ \ref{kvrs} inset. For the case of the spin transport we present 
$T\sigma(\omega)$ in Fig. \ref{svrs} for various moderate-strong 
perturbations. %
\subsection{Lattice size scaling}
\par
Let us now turn our attention to the lattice size scaling. A 
Lorentzian $\kappa(\omega)$ trivially
obeys  a universal $L$ scaling, with the size independent 
quantity to be $\kappa(\omega L)/L$, since $\kappa_{dc}, \tau^{\epsilon}\sim L$, 
Fig.\ \ref{klor}.\cite{PhysRevB.81.205101} %
On the other hand, for strong $J'$, $\kappa(\omega)$ exhibits two $L$-scalings; 
the prominent  impurity contribution at $\omega\sim J'$, which is $\mathcal{O}(1)$, 
does not scale with $L$, Fig.\ \ref{k-ws}(a), while 
 the low frequency  bulk contribution obeys the proposed  scaling for any $J'$.
Moreover, in accord with the evaluation of $J^{*}$  
(Eq.\ (\ref{SumRule})),  Fig.\ \ref{k-ws}(a) shows that the contribution of a 
single impurity dwindles  with respect to the bulk contribution with increasing 
$L$, becoming  negligible in the limit $L\rightarrow\infty$. However, considering the 
thermodynamic limit and a  finite but dilute impurity concentration 
$c_{I}$---which actually  would be a more pragmatic approach to a real 
system---one could plausibly assume that  the high frequency structure 
will be present, and similar scaling   behaviors would hold with the substitution  
$1/L\rightarrow c_{I}$. %
\section{Strong coupling limit}
\begin{figure}
\centering
\includegraphics[scale=0.7]{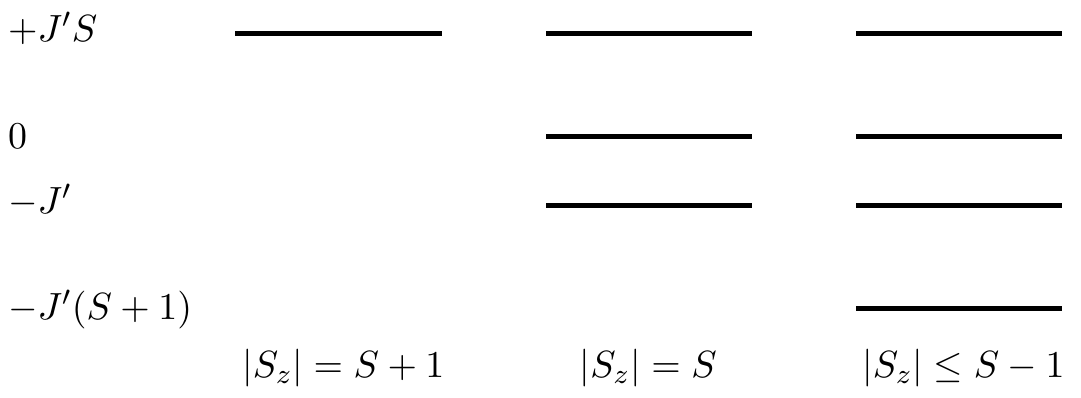}
\caption{The energy spectrum of the isotropic \mbox{3-spin} system, 
Eq.\ (\ref{LocalHamiltonian}).}\label{fig-3} 
\end{figure}
\par
Interesting conclusions can be reached in the strong host-impurity 
limit, $J'\rightarrow\infty$, and are rather useful  to understand 
the high frequency behavior of $\kappa(\omega)$, $\sigma(\omega)$. 
Starting from Eq.\ (\ref{hamiltonian}) for the isotropic point ($\Delta=1$)  
and taking $J=0$, we end up in the 
Hamiltonian of a \mbox{3-spin} system 
\begin{equation}\label{LocalHamiltonian}
\mathcal{H}=J'(\mathbf{s}_{L-1}+\mathbf{s}_{1})\cdot\mathbf{S}\:.
\end{equation}
Exploiting the rotational symmetry and the limited degrees 
of freedom of $\mathcal{H}$,  we  perform an analytical 
diagonalization into different \mbox{$S_{z}$-total} subsectors. 
The lowest/highest $|S_{z}|=S+1$ subsectors are  
$\mathcal{D}=1$ Hilbert spaces. The second lowest/highest  $|S_{z}|=S$ 
subsectors are $\mathcal{D}=3$ Hilbert spaces. The rest $2\seff+1$ $S_{z}$  subsectors, 
with $\seff=S-1$,  are  $\mathcal{D}=4$  Hilbert  spaces. The energy spectrum 
of the \mbox{3-spin} system is shown in  Fig.\ \ref{fig-3}.
\par 
Considering the local energy current operator   
$\tilde{\jen}=J'^{2}\sii\cdot(\sccd_{1}\times\sccd_{L-1})$, we can see that 
its matrix elements vanish 
between non-zero eigenvalues, 
and, consequently, only transitions  between zero and non-zero eigenvalues survive; 
this holds apart from the isotropic point as well.  Particularly, for 
the isotropic Heisenberg model, the only non-vanishing transitions 
correspond to an energy difference $\delta\epsilon=\pm J'$. 
Thus, for $\Delta=1$, $\kappa(\omega)$  
will exhibit only one sharp peak at high frequencies located at 
$\omega\simeq J'$ and it will be independent of $S$. 
Transitions corresponding to $\delta\epsilon=\pm J'$ are between elevated 
eigenstates, Fig.\ \ref{fig-3}, thus the peak at $\omega\simeq J'$  
is expected to diminish  with decreasing temperature  
and finally to vanish  as the system flows towards its ground state. These 
selection rules do not hold for the spin transport where there are more 
allowed transitions. These transitions  involve the ground state $\epsilon_{g}$ 
as well and consequently  resonant peaks  will be present at low $T$. 
Lastly, for  $\Delta\neq1$ there are transitions of  $\tilde{\jen}$ 
which do not  necessarily correspond  to  $\delta\epsilon=\pm J'$ 
and yield a more   complicated high frequency structure for $\kappa(\omega)$. 
%
\begin{figure}
\centering
\includegraphics[scale=0.3]{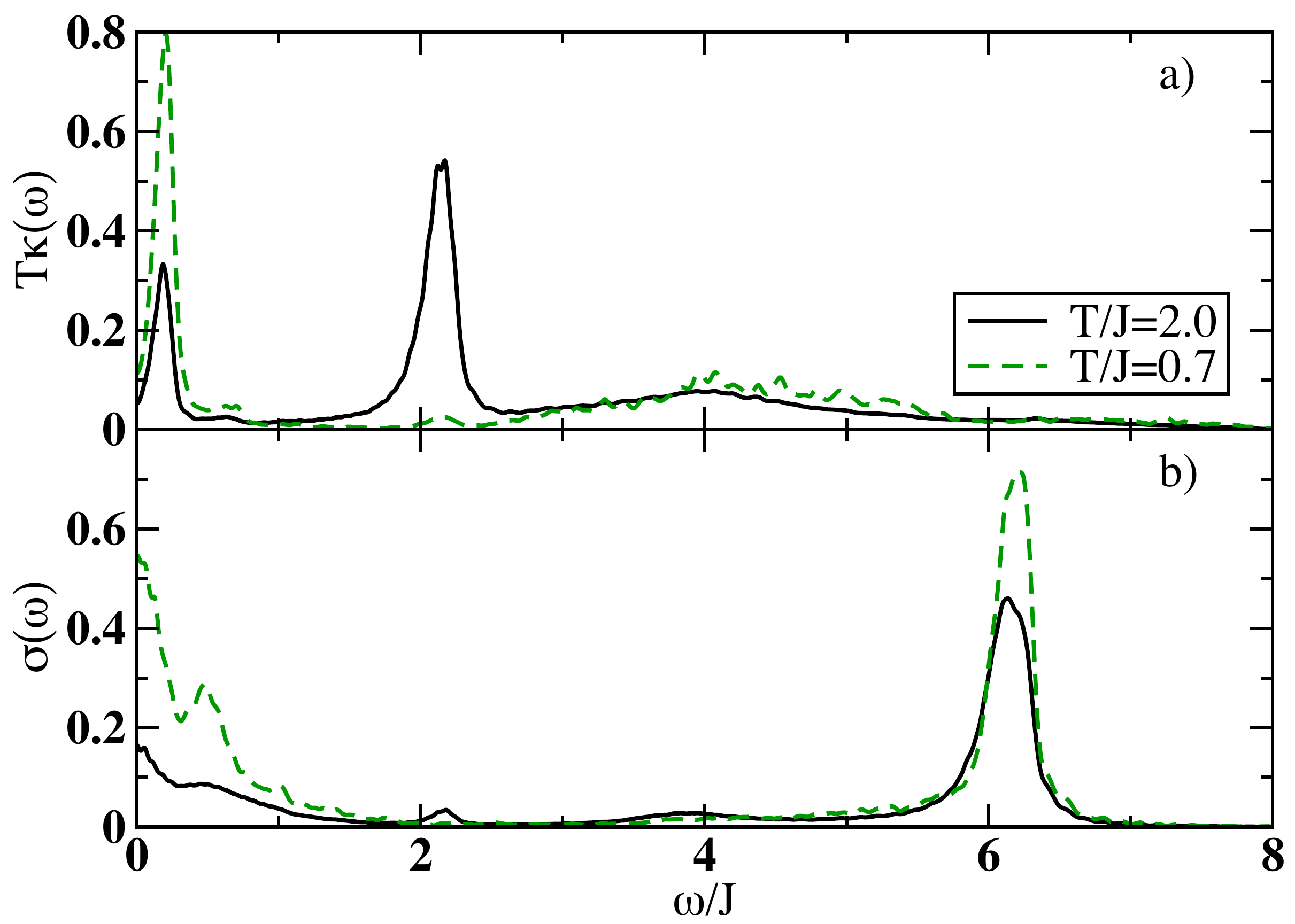}
\caption{(color online).\  a, b) The frequency dependence of 
the thermal conductivity $T\kappa(\omega)$, spin conductivity 
$\sigma(\omega)$ respectively for an $S=2$ impurity  with $J'=2J$, $L=19$, 
$\Delta=1$ and two temperatures $T/J=0.7,2$. Results are obtained via FTLM.}
\label{sk-2t} 
\end{figure}%
\par
The previous analytical arguments are verified in Fig.\ \ref{sk-2t} where 
results are obtained via FTLM for $L=19$ and various perturbations. 
The  frequency dependence of the thermal conductivity 
$T\kappa(\omega)$ Fig.\ \ref{sk-2t}(a) and the spin conductivity 
$\sigma(\omega)$ Fig.\ \ref{sk-2t}(b) is shown, for $S=2$,  $J'=2J$, 
$\Delta=1$ at $T/J=0.7,2$.  
\par
Starting with Fig. \ref{sk-2t}(a) we can see that 
at  high  temperatures the prevalent  contribution to $\kappa(\omega)$ 
comes from the operator $\tilde{\jen}$ yielding  a prominent peak at $\omega=J'$
independent of  $S$ (compare Figs.\ \ref{k-ws}(b) and  \ref{sk-2t}(a)). %
As the temperature decreases  and the system  flows to its ground state 
the peak at $\omega=J'$ decreases  gradually and eventually vanishes. 
Note that the high frequency leap in Fig.\ \ref{sk-2t}(a), present at $T/J=0.7$,  
emerges from  $\propto JJ'$ terms of  the $\jen$ current, Eq. (\ref{encur}). 
In contrast to  the thermal transport,  the local excitations of the spin current in the  
reduced  system, Eq. (\ref{LocalHamiltonian}),  
involve the ground state giving a sharp peak at $\omega=J' (S+1)$ 
which does  not  vanish with decreasing temperature, Fig.\ \ref{sk-2t}(b). 
\begin{figure}
\centering
\includegraphics[scale=0.3]{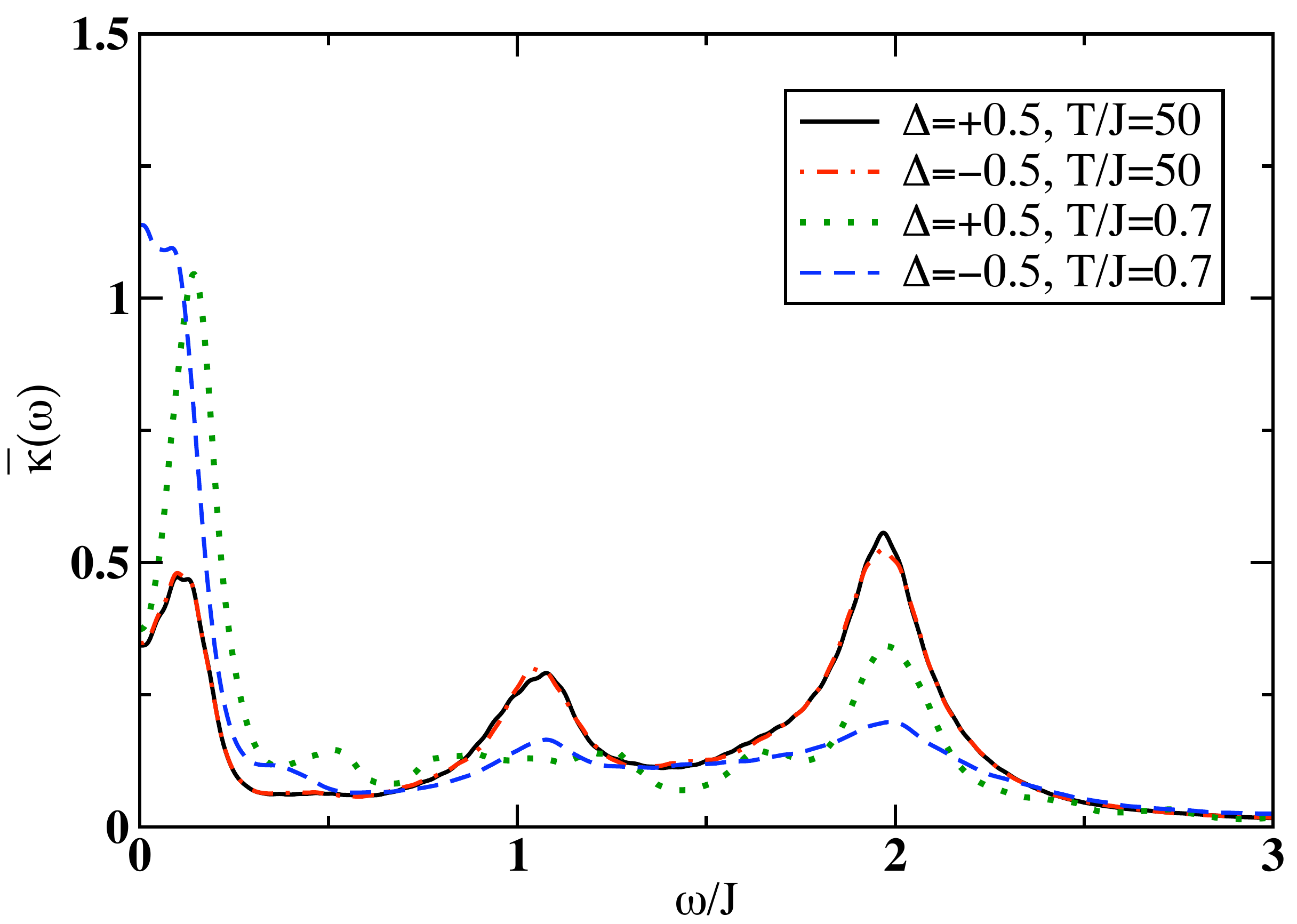}
\caption{(color online).\ 
The frequency 
dependence of the normalized thermal conductivity $\bar{\kappa}(\omega)$ 
for $S=1$, $J'=2J$, $L=19$ and $\Delta=\pm0.5$ at $T/J=0.7,50$. Results
 are obtained via FTLM. %
}\label{ch} 
\end{figure}
\par
In Fig.\ \ref{ch} we present the 
normalized thermal conductivity $\bar{\kappa}(\omega)$ for $S=1$, $J'=2J$, 
$\Delta=\pm0.5$ at $T/J=0.7,50$; results are obtained via FTLM for $L=19$.  
First, in connection with the previous arguments, we can say that the complicated    
high frequency structure of $\kappa(\omega)$  for $\Delta\neq1$ 
indicates that the single excitation at $\omega=J'$  
is  a  unique property of the isotropic Heisenberg model. %
Second, we focus on the low frequency part of 
$\kappa(\omega)$ and particularly on the  stark difference for 
$\Delta\lessgtr0$  at low temperatures. The chain exhibits 
cutting(healing) behavior for $\Delta>0$($\Delta<0$) for an $S=1$ impurity 
embedded in the chain which was previously  reported for an $S=\frac{1}{2}$
 impurity out of the  chain.\cite{PhysRevB.58.5529, PhysRevB.81.205101} %
At high temperatures the curves for $\Delta=\pm0.5$ are one on top of the other. 
As the temperature decreases, $\kappa(\omega)$ for $\Delta=-0.5$ tends to 
obtain a  more Lorentzian-like form, characteristic of the  weak perturbation 
regime. On the contrary, for $\Delta=+0.5$, $\kappa(\omega)$ obtains a 
strongly non-monotonic  behavior with decreasing temperature resembling 
the thermal conductivity of an open chain and signifying  the flow to the strong 
perturbation regime. %
Note that  $\sigma(\omega)$ exhibits a similar low frequency behavior for 
$\Delta=\pm0.5$ which could be plausibly read as evidence of a finite 
$D_S$, yielding a $\sim\delta(\omega)$ contribution to $\sigma'(\omega)$, 
in the pure AHM  for $|\Delta|<1$. %
\begin{figure}
\centering
\includegraphics[scale=0.3]{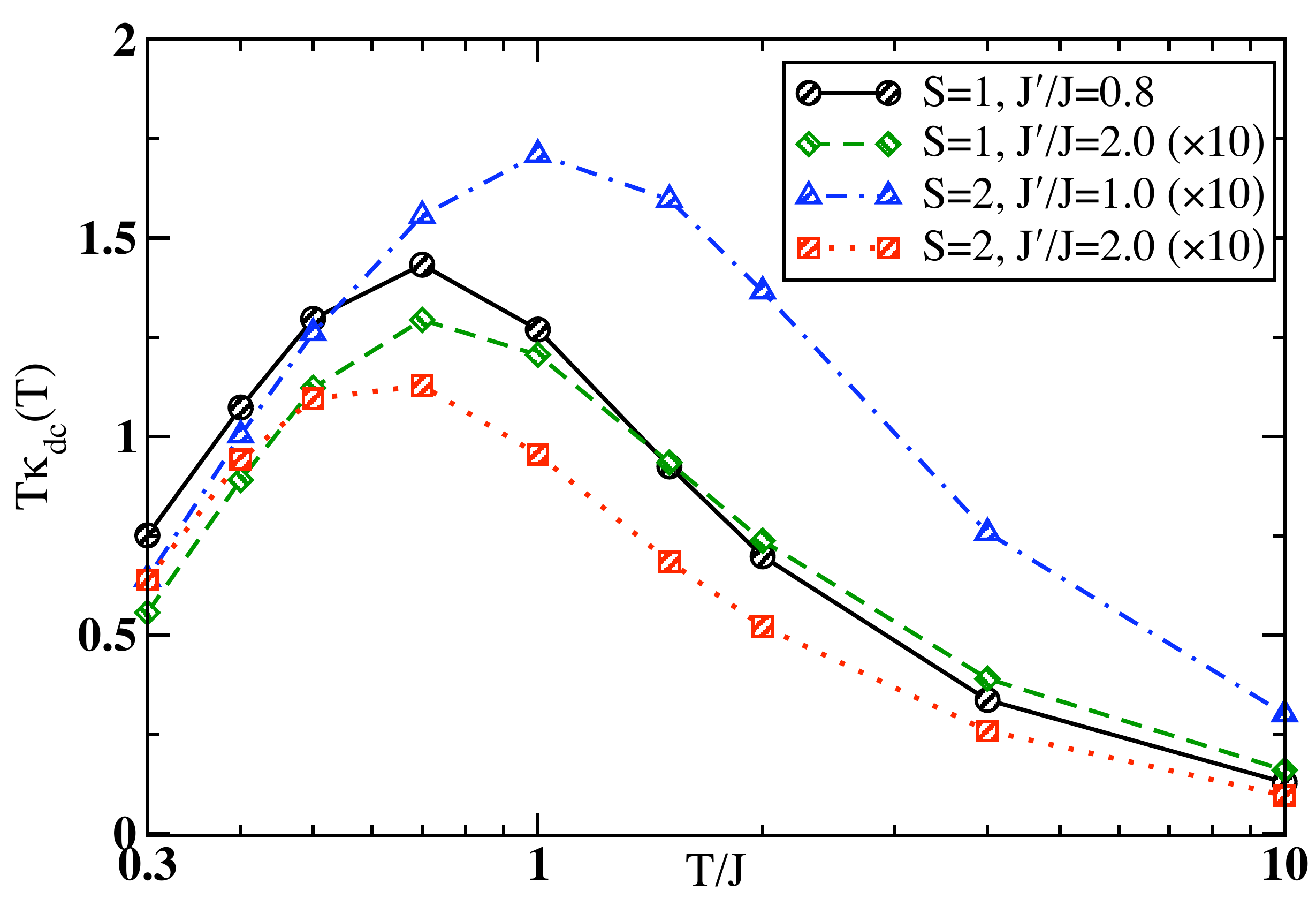}
\caption{(color online).\ 
The temperature 
dependence of  $T\kappa_{dc}(T)$ for $S=1$, $J'/J=0.8, 2$ and 
$S=2, J'/J=1, 2$; results are obtained for $\Delta=1, L=19$ via FTLM. }\label{kdc-vst}
\end{figure} %
\par
It is also interesting to present the temperature dependence of the dc 
value of the thermal conductivity $\kappa_{dc}$ itself since this is a directly 
measurable quantity in experiments. In Fig. \ref{kdc-vst}, $T\kappa_{dc}(T)$ is 
shown for four  characteristic perturbations and $\Delta=1$. 
The main conclusion is that the maximum of $\kappa_{dc}$ occurs at 
different temperatures for different perturbations. Although for $S=1$ 
the host-impurity coupling   quantitatively only  affects   the behavior of 
$\kappa_{dc}$ this is not the case for other impurities like the $S=2$ impurity.
 Generally the maximum of $\kappa_{dc}$ may occur at different 
 temperatures even for the same impurity if this is coupled to the chain with 
 different host-impurity couplings $J'$. 
\par
Finally, we would like to address the screening of the 
impurity by the chain which has been discussed previously in the 
literature.\cite{PhysRevB.46.10866, PhysRevB.64.092410, PhysRevLett.86.516} %
The ground state of the \mbox{3-spin}  system $\epsilon_{g}$ 
is \mbox{$(2\seff+1)$-fold} degenerate, 
Fig.\ \ref{fig-3},    implying  that  the impurity with its neighbors 
form an effective spin $\seffv$  at low  
energies.  
Thus, for a strong $J'$ one can assume that the degrees of freedom of the 
system at low energies will be described by states of the  form 
$|\epsilon_{g}\rangle\otimes|\psi_{L-3}\rangle$, where the 
pseudo spin $\seffv$ couples with the rest of the chain 
with an effective coupling $\jeff$. $\jeff$ can be evaluated considering 
the matrix elements of the operator $h_{1}$ at low 
energies.\cite{PhysRevB.56.654} %
For  an $S=2$ impurity we obtain the effective coupling to be 
{\it ferromagnetic}, $\jeff=-0.25J$, while the larger the $S$ the weaker the 
$\jeff$.  Thus, for a single impurity in a 
finite system, or a finite concentration of impurities in the thermodynamic limit, 
the  picture of the effective spin clearly fails to describe the whole 
frequency range since the weak $\jeff$, $|\jeff|\ll |J'|$,    cannot reproduce 
the high frequency, conspicuous, excitations yielded by a strong  $J'$  
at $\omega\sim J'$.    
\section{Discussion and Conclusions} 
\par
It is worthwhile to devote a few lines to contrast the 
basic points of the present model, where the impurity 
is embedded in the chain (IEC),  
with those of the model studied previously with the impurity located 
out of the chain (IOC).\cite{PhysRevB.81.205101} %
Stark differences arise in the transport properties of the two 
models due to the position of the impurity and the 
way it couples to the pure system. For instance, a weak 
host-impurity coupling $J'$ is a strong perturbation for the 
IEC model, Fig.\ \ref{k-ws}(b), while taking $J'=0$ in the IOC model 
we end up in the pure AHM. Note also that it would not be accurate to perceive 
the difference $|J-J'|$ as a perturbative parameter, simply,  
because the case $J'=J$ does not correspond to the minimal 
perturbation of the heat transport. In addition, 
for different impurities the maximum $\kappa_{dc}$ occurs 
at different $J'$ which does not necessarily correspond to 
the maximum of $\sigma_{dc}$, Figs.\ \ref{ks-dc}, 
\ref{kvrs} inset. %
For spin-$S$ impurities with $S>1$ $\kappa(\omega)$ exhibits a non-monotonic 
form. The absence of a Lorentzian $\kappa(\omega)$  for $S>1$ and any 
host-impurity coupling $J'$ is an indication that $S>1$ impurities constitute 
a strong perturbation for the heat transport of the Heisenberg model. 
This is in sharp contrast to the behavior of 
$\kappa(\omega)$ in the IOC model which obeys a universal  
scaling with the $\mathcal{B}^{2}$ parameter, at least 
for weak-intermediate $J'$.  
\par
Another significant  difference between the two models arises from 
the absence of a $\propto J'^2$ term in the energy current 
of the IOC model. As we have shown in this work, the $\tilde{\jen}$ 
term accounts for the prominent high frequency excitations 
which become the prevalent contribution to $\kappa(\omega)$ for 
strong perturbations. Similarly, the high frequency behavior 
of $\sigma(\omega)$ for strong perturbations is different in 
the IEC and the IOC models due to the absence of a perturbative 
spin current term $\propto J'$ in the latter. From all the above 
one could conclude that the  magnetic impurity in the IEC model 
is a much stronger perturbation for a Heisenberg chain than 
in the IOC model. 
\par
To summarize, studying the thermal $\kappa$ 
and spin $\sigma$ conductivities of the 1D,  \mbox{spin-$\frac{1}{2}$}, AHM 
with an embedded  \mbox{spin-$S$} impurity at finite temperatures $T$, 
we have reached the  following  conclusions: 
\setcounter{conc}{1}(\prconc) %
An $S=1$ impurity  can be considered as a relatively weak perturbation 
for some host-impurity couplings $J'$, in contrast to  $S>1$ impurities 
which have a more drastic effect on $\kappa$. Furthermore, the difference 
between $\sigma_{dc}$ and  $\kappa_{dc}$ is remarkable 
since  the former shows an impressive rigidity  under the influence of the 
impurity. %
\addtocounter{conc}{1} (\prconc) %
$\kappa(\omega)$ obeys a universal scaling $\kappa(\omega L)/L$
for weak and intermediate  $J'$, 
while a strong  $J'$ triggers  the emergence of impurity local excitations 
ruining  the  \mbox{$L$-scaling} of $\kappa(\omega)$.
\addtocounter{conc}{1} (\prconc) %
We have demonstrated  the origin of  these 
resonant modes for a  large $J'$ and a finite system---their position,  
their magnitude, and their  temperature behavior as well, Eq.\ (\ref{SumRule}), 
Fig.\ \ref{sk-2t}.  
\addtocounter{conc}{1} (\prconc) %
The presence of these sharp excitations at low $T$ 
makes the picture of the effective spin $\seff=S-1$ insufficient to describe 
the finite frequency transport properties, Fig.\ \ref{sk-2t}---at least for a single 
impurity in a  finite system or plausibly for a finite concentration of impurities in the 
thermodynamic limit.
\addtocounter{conc}{1} (\prconc) %
Finally, we observe the cutting-healing behavior of the chain 
according to the sign of $\Delta$ as it was previously reported  for magnetic 
impurities out of the chain, 
Fig.\ \ref{ch}.\cite{PhysRevB.58.5529, PhysRevB.81.205101} %
%
\paragraph*{Acknowledgements.}
The author thanks P. Prelov\v{s}ek and X. Zotos for 
their overall contribution  to this paper.  %
This work was supported by  the FP6-032980-2  NOVMAG project. 
%
%
%
%
%
%

\end{document}